# How to fold intricately: using theory and experiments to unravel the properties of knotted proteins

Sophie E. Jackson[1], Antonio Suma[2] and Cristian Micheletti[2]

[1] Department of Chemistry, University of Cambridge, Cambridge CB2 1EW, United Kingdom

[2] SISSA, International School for Advanced Studies, via Bonomea 265, I-34136 Trieste, Italy

**Abstract**

Over the years, advances in experimental and computational methods have helped us to understand the role of thermodynamic, kinetic and active (chaperone-aided) effects in coordinating the folding steps required to achieving a knotted native state. Here, we review such developments by paying particular attention to the complementarity of experimental and computational studies. Key open issues that could be tackled with either or both approaches are finally pointed out.

**Introduction**

Despite the early evidence of a shallowly knotted carbonic anhydrase structure [1, 2], the conviction that proteins had to be knot-free to avoid kinetic traps during folding held until 2000. At that time, a systematic survey [3] of the growing protein databank (PDB) [4] proved unambiguously the occurrence of deeply knotted proteins.

We now know that knotted proteins are uncommon, but not exceptionally rare as they account for about 1% of all PDB entries [5-8]. Their abundance *in vivo* in specific contexts can be highly significant too. For instance, the human ubiquitin C-terminal hydrolase isoform 1 (UCH-L1), that is knotted, accounts for about 2-5% of soluble protein in neurons [9,10].

Both experimental and theoretical approaches have been used to understand the driving forces that coordinate the folding steps leading to knotted native states [6,11-15]. Experiments, unlike present-day simulations, can probe timescales that are sufficiently long to follow the spontaneous folding process. At the same time, folding simulations currently outcompete experiments for the level of detail they can provide of the folding routes. This review aims at conveying such complementarity by focusing on specific aspects that have been tackled with either or both strategies.

## Overview of knotted proteins

An up-to-date non-redundant list of knotted representatives is given in Table I. The range of represented functional families is noticeably broad. Indeed, no common functional rationale for the occurrence of the various knots has been found yet, although knots, which are also known to occur in membrane proteins, could be instrumental to avoid degradation and/or enhance the thermodynamic, kinetic or mechanical stability of proteins [11,16-20]. Indeed, it is intriguing that, in many cases, the knot is close to, or encompasses the active sites of several entangled enzymes, such as ubiquitin carboxy-terminal hydrolyases [5,10,21], RNA methyltransferases [19,22,23], carbamoyltransferases [5], and bacterial phytochromes [24-27].

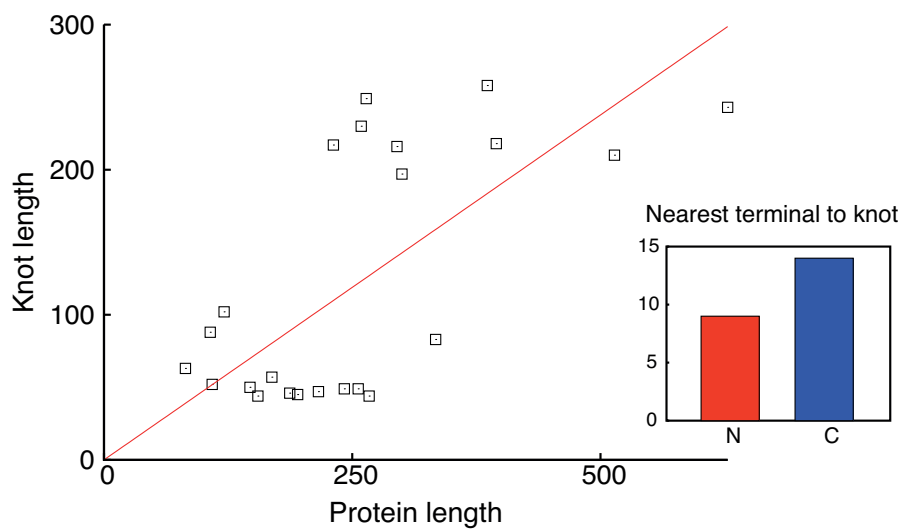

FIGURE 1. **Length and depth of knots in proteins.** The scatter plot presents the knot length *versus* protein length for the 23 minimally redundant representatives in Table I. The Kendall's correlation coefficient, tau = 0.31 and the one-sided *p*-value is 0.018. Of these 23 instances, 9 have the knotted region closest to the N terminus and 14 to the C one, see inset.

| Knot type | Protein or protein function | PDB code | Oligomeric state | Knot length (a.a.) | Chain length (a.a.) | N-terminal knot depth (a.a.) | C-terminal knot depth (a.a.) |
|---|---|---|---|---|---|---|---|
| $3_1$ 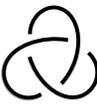 | Carbonic anhydrase | 3MDZ:A | Monomeric | 230 | 259 | 26 | 3 |
| | RNA methyltransferase | 1X7O:A | Homodimeric | 44 | 267 | 190 | 33 |
| | RNA methyltransferase | 4H3Z:B | Homodimeric | 49 | 256 | 89 | 118 |
| | RNA methyltransferase | 4CND:A | Homodimeric | 57 | 169 | 79 | 33 |
| | RNA methyltransferase | 4E8B:A | Homodimeric | 49 | 242 | 164 | 29 |
| | RNA methyltransferase | 3O7B:A | Homodimeric | 47 | 216 | 143 | 26 |
| | RNA methyltransferase | 4JAK:A | Homodimeric | 44 | 155 | 77 | 34 |
| | RNA methyltransferase | 4JWF:A | Monomeric | 46 | 187 | 98 | 43 |
| | RNA methyltransferase | 2QMM:A | Homodimeric | 45 | 195 | 124 | 26 |
| | S-adenosylmethionine synthetase | 4ODJ:A | Homodimeric | 258 | 386 | 16 | 112 |
| | Carbamoyltransferase | 3KZK:A | Homodimeric | 83 | 334 | 169 | 82 |
| | Hypothetical RNA methyltransferase | 1O6D:A | Homodimeric | 50 | 147 | 67 | 30 |
| | Hypothetical protein MJ0366 | 2EFV:A | Homodimeric | 63 | 82 | 10 | 9 |
| | H+/Ca2+ exchanger | 4KPP:A | Monomeric | 218 | 395 | 72 | 105 |
| | Na+/Ca2+ exchanger | 5HWY:A | Monomeric | 197 | 300 | 33 | 70 |
| | N-acetylglucosamine deacetylase | 5BU6:A | Monomeric | 249 | 264 | 10 | 5 |
| | DNA binding protein | 2RH3:A | Monomeric | 102 | 121 | 7 | 12 |
| | DNA binding protein | 4LRV:A | Homo-octameric | 88 | 107 | 8 | 11 |
| | Metal binding protein (zinc-finger) | 2K0A:A | Spliceosome component | 52 | 109 | 21 | 36 |
| $4_1$ 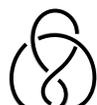 | Bacteriophytochrome | 4GW9:A | Homodimeric | 243 | 628 | 20 | 365 |
| | Ketol-acid reductoisomerase | 1QMG:A | Homodimeric | 210 | 514 | 236 | 68 |
| $5_2$ 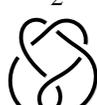 | Ubiquitin carboxy-terminal hydrolase | 2LEN:A | Monomeric or homodimeric | 217 | 231 | 2 | 12 |
| $6_1$ 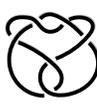 | α-Haloacid dehalogenase I | 3BJX:B | Homodimeric | 216 | 295 | 59 | 20 |

TABLE I. **Representative knotted proteins.** This up-to-date, non-redundant list of knotted representatives is based on a PDB survey specifically carried out for this review. The entries are presented in increasing complexity of the knot and, for each knot type, they are listed in order of decreasing number of represented homologs (number of PDB entries with sequence identity larger than 10%). The oligomeric state is based on the PDB and UNIPROT annotations. Knot types are labelled by the number of crossings in the simplest planar projection of their closed or circularized version (sketched) followed by a conventional indexing subscript. For a stringent determination of physical knots, chains with structural gaps, or with termini not exposed at the protein surface were excluded from survey. Homodimeric entries 3BJX and 4H3Z are accordingly represented by chains B instead of the default chains A, because the latter suffer from the aforementioned issues. The complete list of knotted PDB entries, including a few where knots are likely artefacts due to limited structural resolution, are provided as Supplementary Information. These likely artefactual instances include complex topologies such as $7_7$ or $8_{18}$ knots. Besides these physical knots other forms of protein entanglement have been reported. These include slipknots [14,28], which are observed when a the threading end is folded back onto itself, such that a knot is formed by part of, but not the full-length, chain, and pierced-lasso bundles that are observed when a part of the chain is threaded through a loop formed by a disulphide bond [29].

The physical knots listed in Table I cover four different topologies: the $3_1$, $4_1$, $5_2$ and $6_1$ knots. These are the simplest instances of twist knots that can be tied or untied with a single, suitably chosen, strand threading or passage. Non-twist knots with similar complexity, such as the $5_1$ torus knot, are probably not observed because their folding would be more challenging, requiring at least two strand passages or threading events to be fully tied or untied [6].

A key general question is whether the degree of entanglement observed in proteins differs from that of other compact, globular polymers [30-33]. In this regard, Lua and Grosberg [34] showed that naturally occurring proteins are knotted significantly less than equivalent models of globular homopolymers (which, unlike proteins, lack a defined native state). These results are compatible with the intuition that knots have been selected against in naturally occurring proteins, though not ruled out entirely. Similarly to general polymer models, however, proteins do exhibit a significant correlation between the overall protein length and the length of the knotted region, see Fig. 1. Note that, in the context of a native fold, knots span several tens to hundreds of amino-acid residues and hence largely exceed the minimal length required to tie them. For instance, stretching experiments as well as simulations on a $4_1$-knotted phytochrome both showed that mechanical pulling can reduce the knot size from ~240 to 17 amino-acid residues [25].

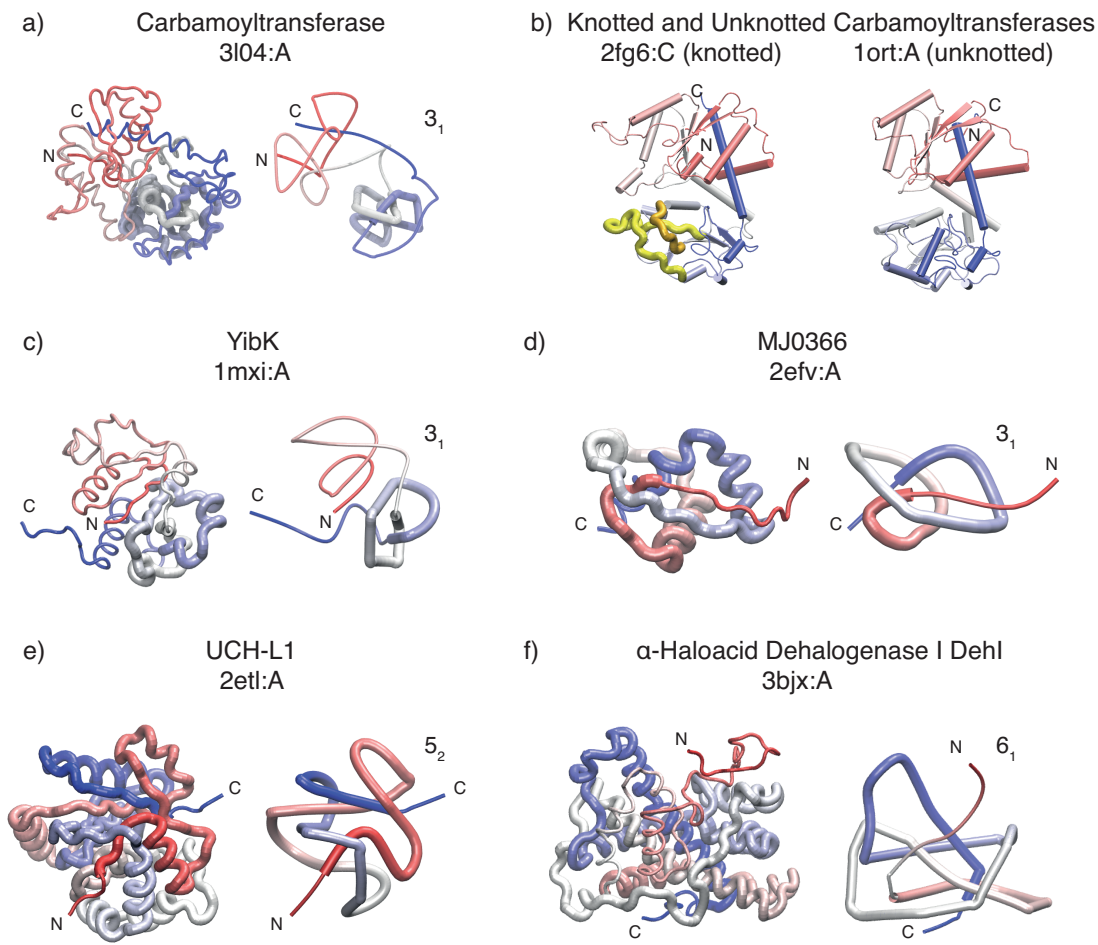

FIGURE 2. **Selected examples of knotted proteins**. In panels a, c-f, a smoothed structural representation is used to highlight the knot. Panel (b) presents a knotted/unknotted pair of carbamoyltransferases [7]. A virtual excision of either or both of the highlighted loops (colored in yellow and orange) unties the knotted variant.

## Trefoil-knotted proteins

**Knotted and unknotted carbamoyltransferases**

A systematic sequence-based comparison of knotted and unknotted proteins [7] showed that knotted carbamoyltransferases, see Fig. 2a, occupy a specific phylogenetic branch off the main trunk of unknotted precursors. Structurally, the key difference is the presence of additional short loops in the knotted variants [5,7], see Fig. 2b. Other types of knotted proteins, including UCH-L1, also have unknotted counterparts differing in the lack of short loops [7]. This suggests that mutation by loop addition may have been a recurrent step in the evolution of knotted proteins from unknotted precursors.

In a series of folding simulations, Skrbic *et al*. [35] took advantage of the sequence and structural similarities of knotted and unknotted carbamoyltransferases to compare their early stages of folding and track differences associated with knotting. For simulations based on a pure Go model, that exclusively rewards native interactions, no propensity to knot formation was observed for either variant. However, when the Go model was complemented with non-native quasi-chemical interactions, a small but systematic propensity to knot (about 0.5%) was found for the natively knotted variant, while it remained negligible for the unknotted one. The knotting events typically involved the threading of the hydrophobic C terminus through loosely structured loop regions.

The involvement of the C terminus in topology-changing events was also seen in Go-model unfolding simulations of unknotted and knotted carbamoyltransferases [17]. In this study knotted variants were found to be more resilient to mechanical unfolding as well.

**Yibk and YbeA**

The bacterial homodimeric proteins YibK and YbeA are, so far, the smallest known members of the α/β-knot methyltransferases (MTases) family [19,36]. Their single-domain monomers are about 160 residues long and accommodate a trefoil knot at a depth of *ca*. 40 residues from the C terminus, see Fig. 2c. They both fold relatively fast from their chemically denatured states [37,38], however, it is now well established that even high concentrations of chemical denaturant do not remove their native $3_1$-knotted topology. This remarkable and unexpected property was proven by trapping the denatured state topology by ligating the protein termini and observing that the resulting cyclized proteins could refold to functional, knotted native states [39]. Later studies on another knotted MTase from *Thermotoga maritima* [40] confirmed that these knotted structures require several weeks under highly denaturing conditions to both unfold and untie themselves [41]. One further property of chemically denatured MTases is that, despite the presence of the knot, which acts as a global constraint, their radii of gyration are similar, and not appreciably smaller than those of denatured, but natively unknotted proteins of similar lengths [42].

Even more remarkable is the fact that both YibK and YbeA are capable of attaining the correct knotted native topology even when molecular plugs are attached to either termini [43]. In the most recent demonstration of this, an *in vitro* transcription-translation system was used to study the *de novo* folding of YibK and YbeA as well as fusion variants obtained by attaching the rapidly folding ThiS domain at either or both termini [44]. The folding rate was slowed down by up to a factor of three when the ThiS plug was attached to the C terminus. No appreciable slowing down was observed when the plug was present solely at the N terminus. The results suggest that the rate-determining folding and knotting events involve the C terminus, close to the location of the knotted region in the native state.

One further aspect relevant for the *in vivo* folding of the knotted methyltransferases is the role of chaperonins. These have been shown to speed up the folding rates of YibK and YbeA by more than an order of magnitude [44,45]. Although the details of the chaperonins' action is still unclear, Jackson and coworkers proposed that they may help unfolding a highly native-like, but unknotted, misfolded state that would otherwise be kinetically trapped. Backtracking from similarly misfolded conformations had previously been observed in simulations [46].

Besides experiments, several folding simulations have been carried out on MTases. The YibK study of Wallin *et al*. [47] was the first to address numerically the folding of a knotted protein. When using a pure Go model, the folding simulations resulted in about 80% formation of native contacts, but not more. This happened because the near-native states were too compact to allow for threading events. However, after introducing non-native attractive interactions between the middle and C-terminal regions of the chain, the knotted native state was reached in all folding attempts [47]. Thus, these non-native interactions were crucial in establishing the correct topology to thread the chain through a loop. Interestingly, two different types of folding routes were observed, one where knotting occurred late and the other early (80% and 20% of native contacts formed, respectively). For the latter set, it would be interesting to verify that the average radius of gyration is comparable with that of still unfolded and unknotted structures, as observed experimentally [42].

In the studies of Sulkowska *et al*. [46] and Prentiss *et al*. [48] pure Go models were used to generate folding trajectories from fully unfolded and untied initial states. They concluded that pure native-centric potentials suffice to drive the folding process towards the lowest-energy, knotted, native state. However, in the Sulkowska study, the yield of successful trajectories was low, around 1-2%, and in the Prentiss study, in order to estimate the fastest speed possible for folding a knotted protein, a minimal and shallowly knotted structure was simulated. In both cases, multiple pathways were observed involving the formation of the knot either at the N or C terminus [46,48] and knotting events occurred either at early [48] or late [46] stages of folding. The knotting modes involved either a direct threading or a slipknotting event. The relative weight of the events depending significantly on the level of structural detail in the model [46].

A recent and interesting twist to the problem was addressed by Cieplak *et al.* [49] who used an optimized Go model to simulate the folding of a nascent chain of YibK. The study showed that the aforementioned low yield of successful folding trajectories could be dramatically enhanced by including co-translational folding effects. In fact, the common knotting event consisted of the formation of a slipknot whilst the C-terminal region of the chain was still attached to the model ribosome, the knot was only able to form in full after the chain was released from it. The model of [49] adopted a tolerant criterion to define native interactions that included the key folding-promoting contacts of Wallin *et al.* [47], and therefore no *ad hoc* non-native contacts were needed to drive the correct folding either co-translationally, or spontaneously. Interestingly, in the latter case, no slipknotting events were observed.

**MJ0366**

The homodimeric protein MJ0366 from *Methanocaldococcus jannaschii* is the smallest known knotted protein. Its monomers are 92 residues in length and feature a shallow trefoil knot at a depth of only 10 residues from the C terminus, see Fig. 2d.

The folding mechanism of MJ0366 has recently been probed experimentally using a number of techniques [50], and evidence found for a highly structured, monomeric, on-pathway intermediate. It is highly likely that this intermediate is not yet knotted as it forms within a millisecond, towards the upper limit of folding rates observed for very small, unknotted proteins. This is followed by a slower second step that involves further folding, and likely knotting, as well as association to form the dimer, similar to results on the folding of $5_2$-knotted UCHs.

The limited protein length and knot depth of MJ0366 has made it the focus of several computational studies. Computational models with different levels of structural detail, force fields and initial conditions have been employed [49,51-53]. In the study of Noel *et al.* [52], based on coarse-grained and atomistic native-centric models, knot formation was observed through both threading and slipknotting mechanisms, the latter being dominant both below the folding temperature and upon extending the C terminus. The study mostly focused on a single monomer of MJ0366 because, within the native-centric scheme used, knotting of the monomers precedes the formation of the dimer [52]. This conclusion, however, is in contrast to the interpretation of experimental NMR HDX data, which indicates that the region of the protein involved in knot formation is not highly structured in the intermediate state [50].

A later study by a Beccara *et al*. [51] used, for the first time, a realistic atomistic force field, *i.e*. non-native centric, to study the folding of a monomer of MJ0366. In this case, computational demand was reduced by using a ratchet-and-pawl scheme to accelerate the evolution of the system to the native state. Only 1% of trajectories successfully reached the knotted native state, knotting occurred *via* direct threading of a loop formed by the earlier formation of the β-sheet, a mechanism later observed in the coarse-grained folding simulations of Najafi and Potestio too [54]. By comparison, slipknotting events were rare and, additionally, a novel mechanism, involving loop-flipping was reported. These studies were later followed by those of refs. [53,55] in which MJ0366 folding was studied from different specific initial conditions. In ref. [53] Noel *et al*. used an unbiased atomistic simulation to study the dynamic evolution from configurations that were unfolded, though already slipknotted. The study of Chwastyk *et al*. [55] instead, used an optimized Go model to study the co-translational folding of MJ0366. They found that the fraction of successful folding trajectories increased dramatically under nascent, co-translational conditions. Knotting events involved direct threading, slipknotting and loop-flipping, similarly to ref. [51] albeit in different proportions, and a further two-loop knotting event was added to the list of mechanisms observed computationally. Interesting, the study reported a reduction in knotting efficiency upon extension of the C terminus, unlike ref. [53] .

**HP0242: a designed trefoil-knotted protein**

In 2010, the Yeates group successfully designed a monomeric $3_1$-knotted protein from the highly entwined homodimer HP0242 [56], thus demonstrating a potential pathway for how knotted proteins might have evolved from unknotted precursors. In experimental folding studies, the designed chain misfolded into a compact, probably unknotted, state before a slow transition, likely involving partial unfolding and knotting, to the knotted native state. Further kinetic studies by the Hsu group suggested that the designed protein folds through multiple intermediates states, only one of which can lead to productive folding [57].

Two computational studies on the knotted HP0242 variant exist. The first study [58], found a surprising lack of deep topological traps using a coarse-grained structure-based model. However, some aspects of the simulations did mimic the experimental results such as the sensitive temperature dependence of successful folding trajectories[58]. The other study [59] used an improved atomic-interaction based coarse-grained model and reported a very high folding success rate, 96%. As had been shown before [58], the folding of the designed knotted protein was considerably slower that its unknotted counterpart. A number of intermediate states were observed including an off-pathway misfolded state that lacked the knot, consistent with the experiments.

**Complex knotted proteins: UCHs and DehI**

The most complex protein knots studied experimentally are from the ubiquitin C-terminal hydrolase (UCH) family, and have a $5_2$ topology. The knot, located only a few residues from the N terminus spans the substrate binding and catalytic site and has been hypothesized to be instrumental in the protein avoiding proteasomal degradation. The folding pathways of two isoforms UCH-L1 (associated with Parkinson's Disease and shown in Fig. 2e) and UCH-L3 have been probed with several techniques and very similar results were obtained.  Refolding after unfolding in chemical denaturants is fully reversible [9]  and proceeds *via* two parallel pathways, each with a metastable intermediate [9,10,60]. The most recent studies of UCH-L1, based on HDX NMR [9], concluded that the two intermediates retain much of the β-sheet core, but differ in the degree to which flanking α-helices are packed against the β-sheet.  Although somewhat circumstantial in nature, the results of this study also suggested that neither intermediate was knotted, consistent with the rate-limiting step being conversion of the intermediate to the native state, and associated with the threading event required to establish the final $5_2$ native topology.

Recently, optical tweezers were used on UCH-L1 to take it to three different unfolded states: unknotted, $3_1$- and $5_2$-knotted and refolding from such states was measured [61].  This study showed that threading to form either a $3_1$- or $5_2$-knotted state slowed folding as expected. However, the inferred energy landscape was much more complex than previously envisioned, as many on- and off-pathway intermediate states were populated during unfolding and refolding. Furthermore, at low/moderate forces the $5_2$-knotted region of the denatured state was much larger than expected, spanning about 40 residues. These results may have implications for the cellular degradation of this class of protein.

The latter problem has been recently tackled computationally in ref. [62].  In this study, the proteasome was simply represented by an effective potential along with constant and periodic pulling forces. Coarse-grained models of a $3_1$-knotted protein were used and the results showed that the knot can hinder or even jam the proteosomal machinery. It is likely that $5_2$-knotted proteins such as UCH-L1, which have much larger twist knots in their denatured states. will have an even greater effect, as observed in the systematic pore translocation computational study of ref. [63].

To our knowledge, no folding simulations have been carried out yet for UCHs. However, Faisca and coworkers have modelled the folding kinetics towards compact structures with a $5_2$ topology on a lattice with pre-assigned geometry and topology [64,65]. They found that target structures with $3_1$ topology were significantly more accessible and stable than $5_2$ ones, that is interesting because both knot types can be tied/untied with a single strand passage.

Native-centric (Go model) folding simulations have nevertheless been carried out for an alpha-haloacid dehalogenase DehI that accommodates a $6_1$ knot at a depth of about 20 residues from the C terminus [8], see Fig. 2f. The analysis of the successful folding trajectories (6 out of 1000) showed that the complex native topology was achieved *via* slipknotting and depended on a large-twisted loop flipping over a smaller twisted loop. The actual occurrence of this knotting mechanism has yet to be established experimentally. The results of recent *in vitro* folding experiments on DehI [66], established that unfolding by chemical denaturants is reversible and proceeds sequentially *via* two intermediates, which differ appreciably in secondary structure content but that are similar in terms of compactness. The authors speculate that knotting may occur before any secondary structure has formed.

**Conclusions**

In our view, the current understanding of knotted proteins could be significantly advanced by tackling the following questions:

Are out-of-equilibrium folding simulations missing any key aspects of the actual process of knot formation? Current computational resources favour the collection of many relatively short, steered, and hence mostly irreversible, folding trajectories over acquiring a few unbiased and necessarily exceedingly long ones. Are there any intrinsically slow spontaneous processes of knot formation that are missed as a result? Is the fact that many trajectories in folding simulations are not successful on the timescales used and do not end up with native knotted structures just a question of computational power? Certainly there are many knotted proteins for which unfolding is fully reversible *in vitro*, however, even for these systems the timescale of folding can be many orders of magnitude slower than the folding of many small, computationally tractable proteins.

What are the roles of native and non-native contacts for the spontaneous folding of knotted proteins? In addition to the native structure, does the primary sequence encode for non-native interactions that are useful to avoid kinetic traps?

Do proteins tied in non-twist or composite knots exist? The spontaneous knotting dynamics of general polymer models indicates that the $5_1$ torus knot and the 'double' trefoil knot, $3_1$-$3_1$ are kinetically accessible and have a lower, but still comparable incidence to $3_1$ and $5_2$ knots [67,68]. Can these knots be observed at all in naturally occurring proteins?

Is it possible to excise a few short loops from a knotted protein to obtain unknotted variants? This fascinating possibility is suggested by structural comparative studies [7], but its viability has not yet been tested experimentally.

Most likely, future breakthroughs in these open issues will be fostered by a tighter integration of the complementary strengths of computational and experimental approaches.

## List of annotated references

Papers of particular interest, published within the period of review, have been highlighted as:
* of special interest
** of outstanding interest

## Acknowledgements
We acknowledge support from the Italian Ministry of Education, grant PRIN 2010HXAW77.

## Appendix A. Supplementary data

Supplementary data associated with this article can be found, in the online version, at http://dx.doi.org/10.1016/j.sbi.2016.10.002.

This experimental work studies how the knotted topology of bacteriophytochrome photoreceptors is maintained in circular permutants.

The study reports on the lack of knots in currently-available RNA structures and contrasts it to the occurrence of knots in proteins.

This work presents a systematic characterization of local and non-local contact propensities in proteins and RNAs and relates it to their apparent different propensity to be knotted.

This computational study explores the principles through which the knot type and location can be encoded by the primary sequence of model knotted proteins.

A coarse-grained model is used to study the extent to which cotranslational folding can assist knot formation.

The first single-molecule study that establishes the complexity of the energy landscape for folding of a knotted protein, as well as the impact of different knotted denatured states on folding. In addition, it characterizes a particularly large $5_2$-knot in the denatured ensemble which has implications for cellular degradation pathways.

A coarse-grained model is used to model the degradation of knotted proteins by pore translocation.

Coarse-grained model simulations are used to study the compliance of different knot types to translocation through a narrow pore. Jamming is observed only at high driving forces and limitedly to twist knots.